\documentclass[fleqn,twoside]{article}
\usepackage{espcrc1}

\usepackage{graphicx}
\usepackage{psfig,epsfig}
\usepackage[figuresright]{rotating}

\newcommand{\AmS}{{\protect\the\textfont2
  A\kern-.1667em\lower.5ex\hbox{M}\kern-.125emS}}

\newcommand{\beq}{\begin{equation}}
\newcommand{\eeq}{\end{equation}}
\newcommand{\bea}{\begin{eqnarray}}
\newcommand{\eea}{\end{eqnarray}}

\newcommand{\Pom}{I\!P}

\newcommand{\FLUKA}{{\sc FLUKA}}
\newcommand{\PEANUT}{{\sc PEANUT}}

\def\dm2{\Delta m^2}
\def\sq2{sin^2(2\Theta)}

\def\aprge{\buildrel > \over {_{\sim}}}   
\def\to{\rightarrow}

\title{\bf Calculation Of Secondary Particles In Atmosphere And Hadronic
Interactions}

\author{
G. Battistoni\address[INFNMi]{INFN and Universit\`a di
Milano,  Dipartimento di Fisica, Milano, 20133, Italy}%
, A. Ferrari\address[CERN]{CERN, Geneva 23, Switzerland},
and P.R. Sala\addressmark[ETH]{ETH, Zurich, Switzerland, on leave of
absence from INFN-Milano}
}       

\begin{document}

\maketitle

\begin{abstract}
Calculation of secondary particles produced by the interaction
of cosmic rays with the nuclei of Earth's atmosphere pose
important requirements to particle production models.
Here we summarize the important features of hadronic simulations,
stressing the importance of the so called ``microscopic'' 
approach, making explicit reference to the case of the \FLUKA{} 
code. Some benchmarks are also presented.
\end{abstract}


\section{Introduction}
Reliable calculations of flux of secondary particles in atmosphere, produced
by the interaction of primary cosmic rays, are essential
for the correct interpretation of the large amount of
experimental data produced by experiments in the field of
astroparticle physics. The increasing accuracy of modern
experiments demands also an improved quality of the calculation tools.
Different ingredients are required to produce
a useful calculation model. Essentially they can be reduced to
three important classes: the primary cosmic ray
spectrum, the modelization of the environment (atmosphere, geomagnetic
field, etc.) and a model of particle production in the hadronic
shower following the collisions of primary c.r.'s with the atmosphere
nuclei. The uncertainty on the primary spectra is dominated by the
systematics of the experiments devoted to their measurement:
in the light of recent measurements by AMS\cite{ams} and BESS\cite{bess},
such an uncertainty is about $\pm$5\% below 100 GeV/nucleon, increasing
to $\pm$10\% at 10 TeV/nucleon.
As far as the environmental description is concerned, 
the large amount of geophysical data now available allows, in principle, 
to achieve a high level of accuracy. 
On the other hand, the knowledge of the features of particle production in 
hadronic interactions is still affected by important 
uncertainties ($\aprge~\pm$15\%). 
Since we have not yet a calculable theory for the
non-perturbative QCD regime, we remain with many different attempts of
building interaction  models, which are tuned by comparison to existing
experimental data. Sometimes, these models can give satisfactory outputs only
in restricted fields of applications.
In this work we discuss in some more detail the situation of these
attempts to describe hadronic interactions, trying to
evidentiate the advantages of the so called ``microscopic'' models, {\it
i.e.} those which try to embed as much as possible of the current
theoretical ideas in terms of elementary constituents and of their
fundamental interactions. We shall make explicit reference to the 
set of models contained in the \FLUKA{} MonteCarlo code\cite{FLUKA1}.

\section{Requirements for Interaction Models}
\label{sec:requirements}

Cosmic ray physics is particularly
demanding from the point of view of particle production models, since
it is in general necessary to consider a wide range of primary energy
and projectiles. 
If we take for example the case of atmospheric neutrinos,
even if we limit ourselves to the class of ``contained'' or ``partially
contained'' events in Super--Kamiokande\cite{superk}, namely events 
with $E_\nu$ in the range from 0.2 to few tens of GeV's, primary
cosmic rays from 1 GeV up to at least 1 TeV per particle must be considered.
Furthermore, primary cosmic rays are composed of protons and nuclei, from
He to Fe (higher mass can be neglected with good approximation).
This is quite a different situation with respect to the standard
case of particle physics, where, in general, almost 
mono-energetic beams are considered. In addition, while in particle physics
the attention is mainly devoted to energy deposition, here instead the
details of single interactions are fundamental to obtain a flux prediction.
As previously stated, it is not possible to rely on a unique model
capable of giving the same quality of results at all energies,
and for all kinematics regimes. 
Therefore particular attention is necessary in order to
assure the right continuity across the transition region between the
different regimes.

Last, but not least, there is the necessity of dealing with different
nuclear species. So far, this problem has been 
mainly solved by recurring to the so called ``superposition'' model, where
a nucleus of mass number A and energy E$_0$ is considered to be equivalent
to A nucleons, each one having energy E$_0$/A.
The question if this is a totally acceptable approximation is still an
argument of discussion. 

\section{Different approaches: parametrized vs ``microscopic'' models }
\label{sec:appro}
 
In building a suitable model for particle production, 
we can identify two main 
different attitudes: parametrized codes and ``microscopic'' models. 
In the first case, one 
relies upon analytical formulas derived from some general
phenomenological features of particle production, and
with parameters that are obtained from fits to experimental data.
An example of these fundamental properties of hadronic interactions, is
Feymann scaling, which is known to be a rather good (although not
completely exact)
approximation, 
especially in the forward (``fragmentation'') kinematic region which is
dominant in secondary production by cosmic rays. This
property can be expressed as follows. The number of  pions of energy $E_\pi$
produced in an interaction  by a primary proton of  energy $E_0$
is well  represented  expression:
\begin{equation}
 {dn_\pi \over dE_\pi} (E_\pi, ~E_0)  \simeq {1 \over E_\pi }
~F \left ( {E_\pi \over E_0} \right )  
\label{eq:scaling}
\end{equation}
where the function $F(x)$,
$x$ = $E_\pi/E_0~\sim~x_{Feynman} $,
 is approximately independent
from the primary energy  $E_0$,  and decreases monotonically
from a finite value for $x \to 0$,   to zero for $x\to 1$.
The shape of the curve can be easily expressed by means of a
combination of elementary functions with just a few parameters that can be extrated from
experimental data sets. 
A noticeable example of this
kind of approach is at the basis of the work which has been carried on
by the  Bartol group for many
years, producing many valuable results, and in particular 
the prediction for atmospheric neutrino fluxes\cite{bartol}. 
For this purpose, they constructed
the TARGET numerical model, a module which can be easily inserted in
any cascade program. 
TARGET considers hadron interactions on light nuclei, like
Oxygen and Nitrogen, subdividing the available energy between
leading nucleons and other produced hadrons on the basis of an
assumed elasticity function. Pions, kaons, etc. are then produced according
to parametric formulas reproducing the scaling properties described above.
Experimental data at different energies fix the parameters and guide the
evolution of multiplicities as a function of energy. 
At low and intermediate energies, resonance production is considered.
Care has been taken to assure event by event energy conservation.
The advantage of this kind of approach, mainly used in the framework of a
1-dimensional description\footnote{an upgrade of the TARGET model, also in view
of 3--D applications,  has been presented in \cite{target}},  is that it
can lead to the comprehension of some important and general
properties  of particle production in terms of analytical
expressions. The price to pay is the
lack of generality and  the spoiling of correlations among reaction products.

The second line of approach is instead the use of 
models which try to describe interactions in terms of the properties of
elementary constituents. 
In principle one would like to derive all features of ``soft" 
interactions (low-$p_T$ interactions) from the QCD Lagrangian, as it is 
done for hard processes. Unfortunately the large value taken by the 
running coupling constant prevents the use of perturbation theory. 
Indeed, in QCD, the color field acting among quarks is carried by the
vector bosons of the strong interaction, the gluons, which are
``colored" themselves. Therefore the characteristic feature of gluons
(and QCD) is their strong self-interaction. If
we imagine that quarks are held together by color lines of force, the
gluon-gluon interaction will pull them together into the form of a tube
or a string. Since quarks are confined, the energy required to
``stretch" such a string is increasingly large until it suffices to
materialize a quark-antiquark couple from the vacuum and the string
breaks into two shorter ones, with still quarks at both ends.
Therefore it is not unnatural that because of quark confinement, theories
based on interacting strings emerged as a powerful tool in understanding
QCD at the soft hadronic scale (the non-perturbative regime).
Different implementations of this idea exist, having obtained
remarkable success in describing the features of hadronic interactions.
Some of these codes have already found applications in astroparticle
physics. We can quote a few major examples:
\begin{enumerate}
\item the atmospheric neutrino flux by M. Honda et
al.\cite{hkkm}, which has been obtained by a combinations of 
microscopic codes embedded into an original shower code;

\item the CORSIKA shower code\cite{corsika}, which offer the choice
among different microscopic models.

\item the already mentioned general purpose MonteCarlo code \FLUKA{}, which is now
applied also to cosmic ray physics by us and other authors, and that will
be later described in more detail.
\end{enumerate}

In our opinion, in the microscopic approach 
each step has sound physical basis and allows
to reach a deep understanding of the phenomena and a high reliability of
predictions.
 The performances are optimized comparing with particle production data at
single interaction level. 
The final predictions are obtained with a minimal set of free parameters,
fixed for all energies and target/projectile combinations.
Results in complex cases as well as scaling laws and properties come out
naturally from the underlying physical models. The basic conservation
laws are fulfilled ``a priori''.
A microscopic model can reach a very high level of
detail, at least in principle,
and therefore is a good choice when aiming at precision calculations.
The price to pay is the loss of simplicity and flexibility: there are no
more simple analytical guidelines which allow to understand the
basic properties. Furthermore microscopic codes are more demanding than
parametrizations in terms of computing power.
Parametrized models (if parametrizations are performed at the
level of single interactions) are instead useful as a first, fast and flexible
approach.  

Instead, models tuned on ``integral data'', like
calorimeter resolutions, thick target yields etc.,
can be very inaccurate at the level of single
interactions, as shown in ref.~\cite{carminati} for the case of
{\sl GEANT-GHEISHA}: such a model cannot be used to obtain a reliable
calculation of particle fluxes. In our opinion this might be a problem
for the low energy ($<$80 GeV) calculations performed with CORSIKA,
when {\sl GHEISHA} is selcted under that energy threshold.

In the following section
we shall concentrate on the example of the \FLUKA{} MonteCarlo code.

\section{The FLUKA model}
\label{sec:fluka}

The modern \FLUKA{}\cite{FLUKA1} is an interaction and transport 
MonteCarlo code able to treat with a high degree of detail 
the following problems:

\begin{itemize}
\item Hadron-hadron and hadron-nucleus interactions 0-100~TeV
\item Electromagnetic and $\mu$ interactions 1~keV-100~TeV
\item Charged particle transport - ionization energy loss 
\item Neutron multigroup transport and interactions 0-20~MeV
\item Nucleus-nucleus and hadron-nucleus interactions 0-10000~TeV/n:
      {\it under development}
\end{itemize}

Here we shall review the
two hadronic models which are used inside \FLUKA{} to describe nonelastic
interactions:
\begin{itemize}
\item The ``low-intermediate'' energy one, \PEANUT{}, which covers the energy 
range up to 5~GeV
\item The high energy one which can be used up to several tens of TeV,
based on the color strings concepts sketched in the previous section.
\end{itemize}

The nuclear physics embedded in the two models is very much the same.
The main differences are a coarser nuclear description (and no preequilibrium
stage) and the Gribov-Glauber cascade for the
high energy one.

\subsection{The \PEANUT{} Model}

Hadron-nucleus non-elastic interactions are often described
in the framework of the IntraNuclear Cascade (INC) models.
This kind of model was 
developed at the very beginning of the history of energetic nuclear interaction modelling, but it 
is still valid and in some energy range it is the only 
available choice.
Classical INC codes were based on a more or less accurate treatment of 
hadron multiple collision processes in nuclei, the target being assumed 
to be a cold Fermi gas of nucleons in their potential well.
The hadron-nucleon cross sections used in the calculations are 
free hadron--nucleon cross 
sections. Usually, the only quantum mechanical concept incorporated was 
the Pauli principle. Possible hadrons were often limited to pions and 
nucleons, pions being also produce or absorbed via isobar (mainly 
$\Delta_{33}$) formation, decay, and capture. 
Most of the historical weaknesses of INC codes have been mitigated or 
even completely solved in some of the most recent 
developments~\cite{FLUKA1,MASHNIK}, thanks to the inclusion of a 
so called ``preequilibrium'' stage, and to further quantistic effects including
coherence and multibody effects. 

All these improvements are considered in the \PEANUT{} 
(PreEquilibrium Approach to NUclear Thermalization) model of FLUKA.
Here the reaction mechanism is modelled in by explicit
intranuclear cascade smoothly joined to statistical (exciton)
preequilibrium emission~\cite{Gad92} and followed by evaporation 
(or fission or Fermi break-up) and gamma deexcitation.
In both stages, INC and exciton, the nucleus is modelled as a sphere
with density given by a symmetrized Woods-Saxon~\cite{Gry91}
shape with parameters according to the droplet model~\cite{Myers} for A$>$16, 
and by a harmonic oscillator shell model for light
isotopes (see~\cite{Elton}). 
The effects of the nuclear and Coulomb potentials outside the nuclear 
boundary are included. Proton and neutron densities are generally different.
Binding Energies are obtained from mass tables.
Relativistic kinematics is applied at all stages, with
accurate conservation of energy and momentum including those of the residual
nucleus. Further details and validations can be found
in~\cite{FLUKA1}.

For energies in excess of few hundreds MeV the inelastic channels
(pion production channels) start to play a major role.
The isobar model easily accommodates multiple pion production, for 
example allowing the presence of more than one resonance in the 
intermediate state (double pion production opens already at 600~MeV 
in nucleon-nucleon reactions, and at about 350~MeV in pion-nucleon ones).
Resonances which appear in the intermediate states can be treated as real 
particles, that is, they can be transported and then 
transformed into secondaries according to their lifetimes and decay 
branching ratios. 

\subsection{The Dual Parton Model for high energy}

A theory of interacting strings can be managed by means of 
the Reggeon-Pomeron calculus in the 
framework of perturbative Reggeon Field Theory\cite{Collins},
an expansion already developed before the 
establishment of QCD.
Regge theory makes use explicitly of the constraints of analyticity and 
duality. 
On the basis of these concepts, calculable models can be constructed
and one of the most successful attempts in this field is the so called 
``Dual Parton Model'' (DPM), originally developed in Orsay in
1979~\cite{DPMORI}.  
It provides the theoretical framework 
to describe hadron-nucleon interaction from several
GeV onwards. 
In DPM a hadron is a low-lying excitation of an open
string with quarks, antiquarks or diquarks sitting at its ends. In
particular mesons  are
described as strings with their valence quark and antiquark at the
ends. (Anti)baryons are treated like open strings with a (anti)quark and a
(anti)diquark at the ends, made up with their valence quarks.

\begin{figure}[hbtp]
\begin{center}
\makebox[80mm]{\psfig{file=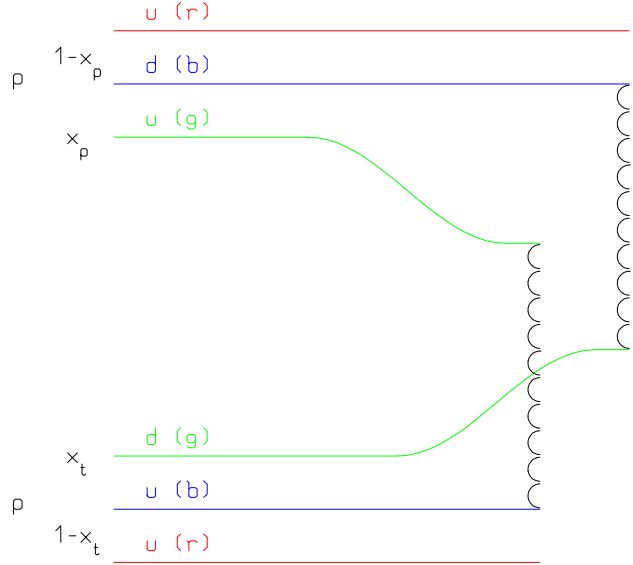,width=80mm%
,bbllx=60pt,bblly=60pt,bburx=550pt,bbury=540pt}}
\caption{Leading two-chain diagram in DPM for $p-p$ scattering. The color
(red, blue, and green) and quark combination shown in the figure is just 
one of the allowed possibilities.\label{fig:ppchain} }
\end{center}
\end{figure}

At sufficiently high energies,
the leading term in high energy scattering 
corresponds to a ``Pomeron'' ($\Pom$) exchange (a closed string exchange
with the quantum numbers of vacuum), 
which has a cylinder topology. By means of the optical theorem, connecting
the forward elastic scattering amplitude to the total inelastic cross
section, it can be shown that 
from the Pomeron topology it follows that two hadronic 
chains are left as the sources of particle production
(unitarity cut of the Pomeron). 
While the partons 
(quarks or diquarks) out of which chains are stretched carry a net 
color, the chains themselves are built in such a way to carry no net 
color, or to be more exact to constitute color singlets like all 
naturally occuring hadrons. In practice, as a consequence of color
exchange in the interaction, each colliding hadron splits into two
colored system, one carrying color charge $c$ and the other $\bar c$.
These two systems carry together the whole momentum of the hadron. The
system with color charge $c$ ($\bar c$) of one hadron combines with the
system of complementary color of the other hadron, in such a way to
form two color neutral chains. These chains appear as two back-to-back
jets in their own centre-of-mass systems.
The exact way of building up these chains depends on the nature of the
projectile-target combination (baryon-baryon, meson-baryon,
antibaryon-baryon, meson-meson). Let us take as example the
case of nucleon-nucleon (baryon-baryon) scattering.
In this case, indicating with $q^v_p$ the valence quarks of the projectile, 
and with $q^v_t$ those of the target, and assuming that the quarks sitting 
at one end of the baryon strings carry momentum fraction 
$x^v_p$ and $x^v_t$ respectively, the resulting chains are
$q^v_t-q^v_p q^v_p$ and $q^v_p-q^v_t q^v_t$, as shown in
fig.~\ref{fig:ppchain}.

Energy and momentum in the centre-of-mass system of 
the collision, as well as the invariant mass squared of 
the two chains, can be obtained from:
\bea
E^*_{ch1} & \approx & \frac{\sqrt{s}}{2} ( 1 - x^v_p + x^v_t ) \nonumber \\
E^*_{ch2} & \approx & \frac{\sqrt{s}}{2} ( 1 - x^v_t + x^v_p ) \nonumber \\
p^*_{ch1} & \approx & \frac{\sqrt{s}}{2} ( 1 - x^v_p - x^v_t ) =
- p^*_{ch2} \label{eq:xeps} \\
s_{ch1} & \approx & s ( 1 - x^v_p ) x^v_t \nonumber \\
s_{ch2} & \approx & s ( 1 - x^v_t ) x^v_p \nonumber
\eea

The single Pomeron exchange diagram is the dominant contribution,
however higher order contributions with multi-Pomeron exchanges become
important at energies in excess of 1~TeV in the laboratory. They
correspond to more complicated topologies, and DPM provides a way for
evaluating the weight of each, keeping into account the unitarity
constraint. Every extra Pomeron exchanged gives rise to two
extra chains which are built using two $q\bar q$ couples excited from
the projectile and target hadron sea respectively. The inclusion of 
these higher order diagrams is usually referred  to as {\it multiple 
soft collisions}.

Two more ingredients are required to completely settle the problem. The 
former is the momentum distribution for the $x$ variables of valence 
and sea quarks. Despite the exact form of the momentum distribution
function, $P(x_1,..,x_n)$, is not known, general considerations based on
Regge arguments allow to predict the asymptotic behavior of this 
distribution whenever each of its arguments goes to zero. The behavior 
turns out to be singular in all cases, but for the diquarks. A 
reasonable assumption, always made in practice, is therefore to 
approximate the true unknown distribution function with the product of 
all these asymptotic behaviors, treating all the rest as a 
normalization constant.

The latter ingredient is a hadronization model, which must 
take care of transforming each chain into a sequence of physical 
hadrons, stable ones or resonances. The basic assumption is that of
{\it chain universality}, which assumes that once the chain ends and the 
invariant mass of the chain are given, the hadronization properties are 
the same regardless of the physical process which originated the chain.
Therefore the knowledge coming from hard processes and $e^+e^-$ 
collisions about hadronization can be used to fulfill this task. There 
are many more or less phenomenological models which have been developed 
to describe hadronization (examples can be found 
in~\cite{JETSET,BAMJET}). In principle hadronization properties too can 
be derived from Regge formalism~\cite{Kaifrag1}.

%
It is possible to extend DPM to hadron-nucleus collisions 
too~\cite{DPMORI}, making use of the so called Glauber-Gribov approach. 
Furthermore DPM provides a theoretical framework for
describing hadron diffractive scattering both in hadron-hadron and 
hadron-nucleus collisions. General informations on diffraction in DPM
can be found in~\cite{GOULIA} and details as well as practical 
implementations in the DPM framework in~\cite{HANDIF1}.

At very high energies, those of interest for high energy cosmic ray studies 
(10--10$^5$~TeV in the lab), hard processes cannot be longer 
ignored. They are calculable by means of perturbative QCD and
can be included in DPM through proper unitarization 
schemes which consistently treat soft and hard processes together.
The interested reader can find more informations as well as practical 
implementations and results in~\cite{DPMORI,HANNES1}.

DPM exhibited remarkable successes in 
predicting experimental observables. The quoted references include a 
vast amount of material showing the capabilities of the model when 
compared with experimental data. However,
it must be stressed that other models are available, but most of them share an 
approach based on string formation and decay. For example, the {\it 
Quark Gluon String Model}~\cite{QGSM} has been developed more or less in
parallel with DPM. This model shares most of the basic features of DPM, 
while differing for some details in the way chains are created and in
the momentum distribution functions.

\section{Benchmarks of the \FLUKA{} Model}
\label{sec:bench}

The predictions of \FLUKA{} have been checked with a large set of
experimental data collected in accelerator experiments. 
Here we shall limit ourselves to show only a few
examples, among the most important in view of the application of
the code to cosmic ray applications.

\begin{figure}[htb]
\begin{center}
\begin{tabular}{cc}
\makebox[75mm]{\psfig{file=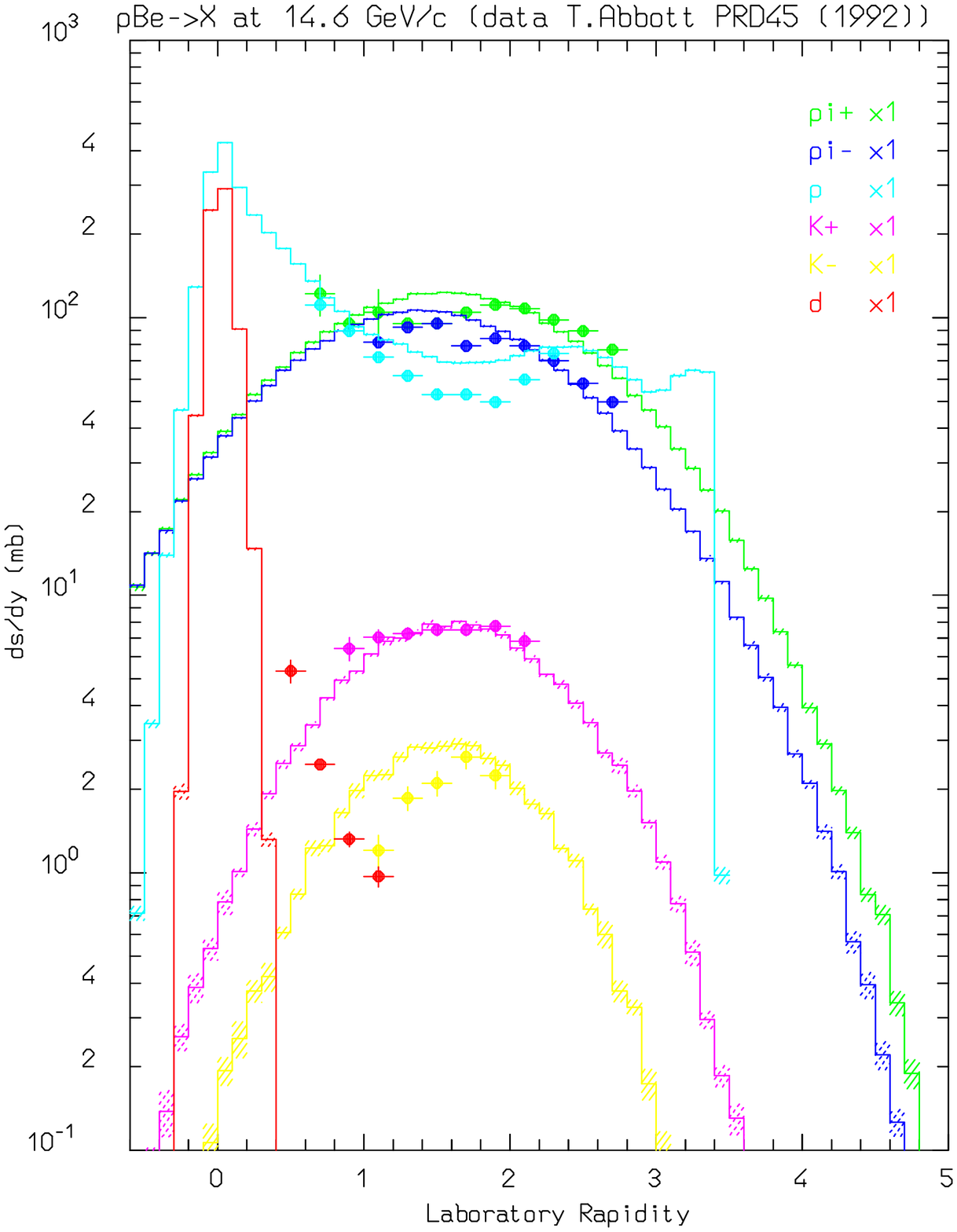,width=75mm%
,bbllx=60pt,bblly=100pt,bburx=590pt,bbury=740pt}} &
\makebox[75mm]{\psfig{file=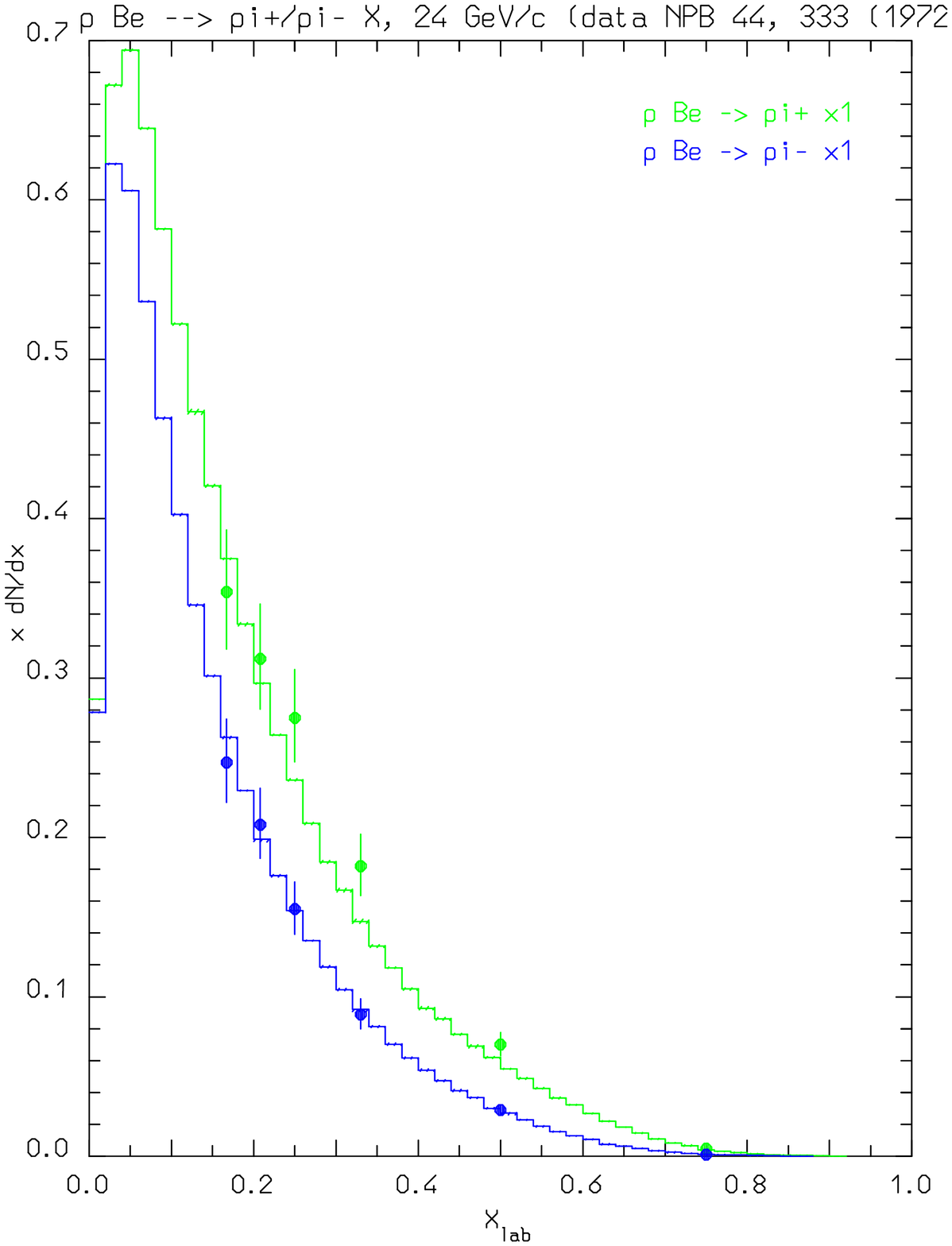,width=75mm%
,bbllx=60pt,bblly=100pt,bburx=590pt,bbury=740pt}} \\
\end{tabular}
\caption{
Rapidity distribution of $pi^{+/-}$ and K$^{+/-}$
for 14.6~GeV/c protons on Be (left, data from ref.~\protect\cite{abbott}),
and $X_{lab}$ distribution for $pi^{+/-}$ for 24~GeV/c 
protons on Be (right, symbols extrapolated from the double differential
cross section reported in ref.~\protect\cite{eichten}).
Histograms are simulation results.
\label{fig:pbey}
}	
\end{center}
\end{figure}

Two sets of data are of particular relevance to check the quality
of a model to be used for the calculation of atmospheric neutrino
fluxes.  These concern p-Be collisions and are reported in
fig.~\ref{fig:pbey}: in ref.\cite{abbott} the central rapidity region has been
mainly explored, while
in ref.\cite{eichten} the forward region has been investigated. 
In both cases the agreement of \FLUKA{} predictions is
quite good.

\begin{figure}[ht]
\begin{center}
\begin{tabular}{cc}
\makebox[75mm]{\psfig{file=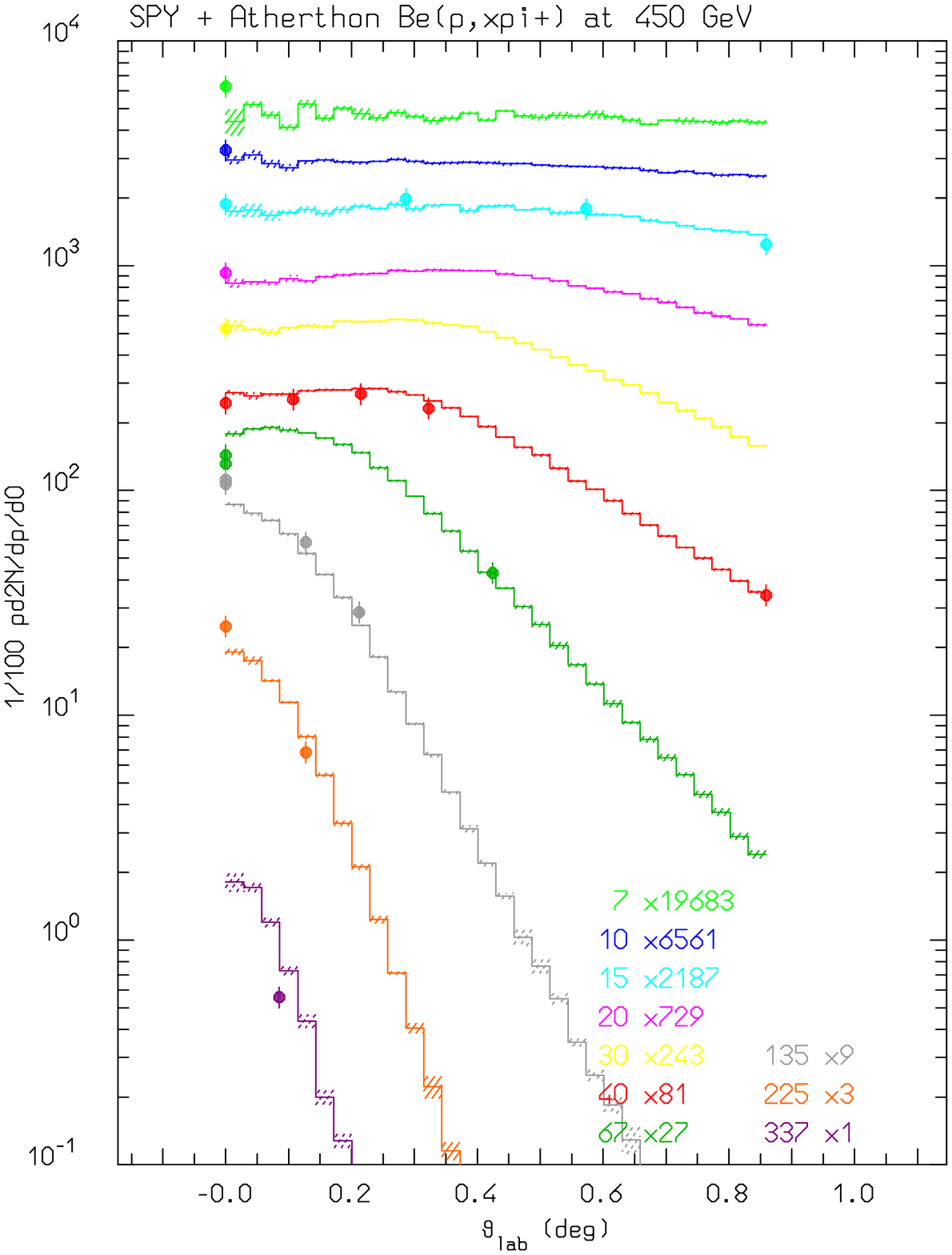,width=75mm%
,bbllx=60pt,bblly=100pt,bburx=590pt,bbury=740pt}} &
\makebox[75mm]{\psfig{file=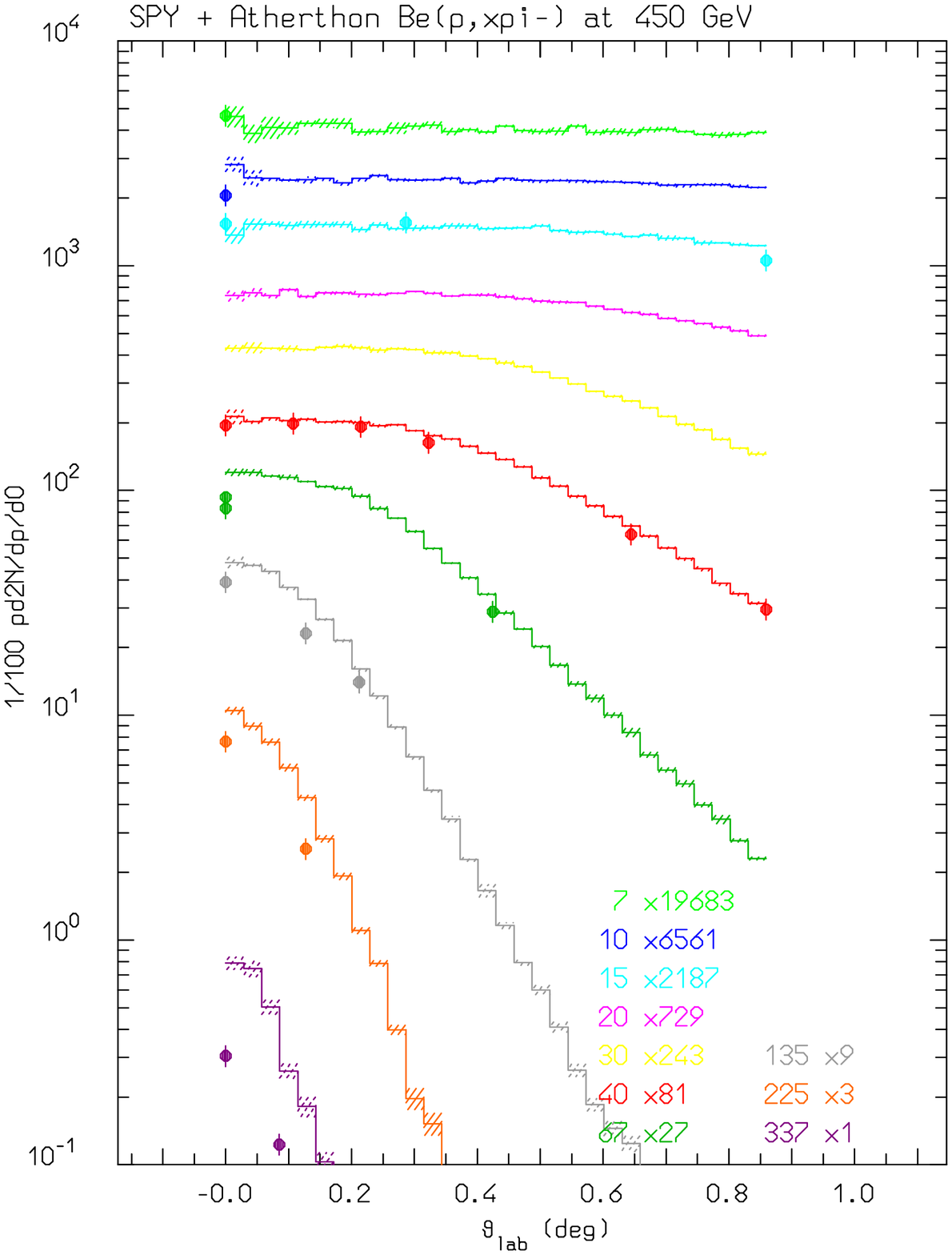,width=75mm%
,bbllx=60pt,bblly=100pt,bburx=590pt,bbury=740pt}} \\
\end{tabular}
\caption{
Double differential cross section for $pi^+$ (left) and $\pi^-$ (right)
production for 450~GeV/c protons on a 10 cm thick Be target 
(data from ref.~\protect\cite{atherton} and \protect\cite{spy}).
Data are given as a function of $\theta_{lab}$ and for different
momentum bins. From top to bottom: 7, 15, 40 and 135 GeV/c, scaled
respectively by a factor of 19683, 2187, 81 and 9.
Histograms are simulation results.\label{fig:spypi}
}
\end{center}
\end{figure}

 Measurements of  $\pi^{\pm}$ and $K^{\pm}$ production rates by 400 GeV/c
 protons on Be targets were performed by Atherton et al. \cite{atherton}
 for secondary particle momenta above 60 GeV/c and up to 500 MeV/c of
 transverse momentum.
 Recently the NA56/SPY (Secondary Particle Yields) experiment  \cite{spy} 
 was  devoted to directly measure these yields in the momentum
 region below 60 GeV/c. The SPY experiment measured
 the production at different angles $\theta$ and momenta $P \leq 135$ GeV/c
 down  to 7 GeV/c for pions, kaons, protons and their antiparticles,
 using a 450 GeV/c proton beam impinging on Be targets.
 These data were extremely valuable to improve the hadronization model
 of \FLUKA{} so to arrive at the present version.
 \FLUKA{} is in agreement with the Atherton and the SPY  measurements
 at the level of $\sim 20 \%$  in the whole momentum range of all secondaries,
 with the exception of a few points mostly for negative kaons.
 The case of pions is reported in fig.\ref{fig:spypi}.
 Also the $\theta$ dependence of the measured yields  
 is reasonably described by \FLUKA{}.
 The measured $K^{\pm}/\pi^{\pm}$ ratios are reproduced to
 better than $20 \%$ below 120 GeV/c.

This example was of particular relevance, since other attempts, using
for instance the hadronic interfaces of {\sl GEANT} (and in particular
{\sl GEANT-GHEISHA}) yielded a much worser agreement, as shown in
ref.\cite{now98}.

\section{Example of calculations of particles in atmosphere}
\label{sec:example}

In the last years, the \FLUKA{} interaction models has been used to
produce new predictions for the 
atmospheric neutrino fluxes within a
full 3D calculation\cite{flukanu}. These fluxes have been also considered
by the Super--Kamiokande experiment\cite{superk}. In the framework of
the same group, a new calculation has been developed, choosing again
another microscopic code based on the Dual Parton Model:
DPMJET-III\cite{dpmjet3,hkkm2}. It gives results close to those
obtained with \FLUKA{}.

In the last two years a considerable amount of work has been devoted 
to cross check the validity of
the calculation model. As far as the \FLUKA{} approach is concerned, 
at least two remarkable results can be quoted:
\begin{enumerate}
\item The reproduction of the features of primary proton flux as a function
of geomagnetic latitude as measured by AMS\cite{zuccon}, thus showing that 
the geomagnetic effects and the overall geometrical description of the
3--D setup are well under control. 
In addition, the same work shows that
also the fluxes of secondary $e^+e^-$ measured at high altitude (eventually
the last stage of the chain decay of produced mesons) are
reproduced.

\item The good reproduction of the data on muons in atmosphere as measured
by the CAPRICE experiment\cite{caprice}, both at ground level and at different
floating altitudes\cite{caprice_fluka}, when starting from the same primary
flux (Bartol fit) used to generate atmospheric neutrinos. See the quoted
reference for relevant plots and numbers.


The fluxes of atmospheric muons are strictly related to the neutrino ones,
because almost all $\nu$'s are produced either in association, with, or in the
decay of $\mu^\pm$. Therefore it is possible to conclude that, for that
choice of primary spectrum, the $\nu$ fluxes predicted by \FLUKA{} are probably
in the right range.
To a large extent the agreement between the original HKKM\cite{hkkm} and Bartol\cite{bartol}
calculations of the $\nu$ fluxes, despite they started from different estimates
of the primary flux and different hadronic interaction models, is
not casual, but the result of the $\mu$ constraint.
Furthermore, the agreement exhibited by the \FLUKA{} simulation for muons
of both charges gives confidence on
the predictions of FLUKA for the parent mesons of muons (mostly pions).
\end{enumerate}

\begin{figure}
\begin{center}
\makebox[100mm]{\psfig{file=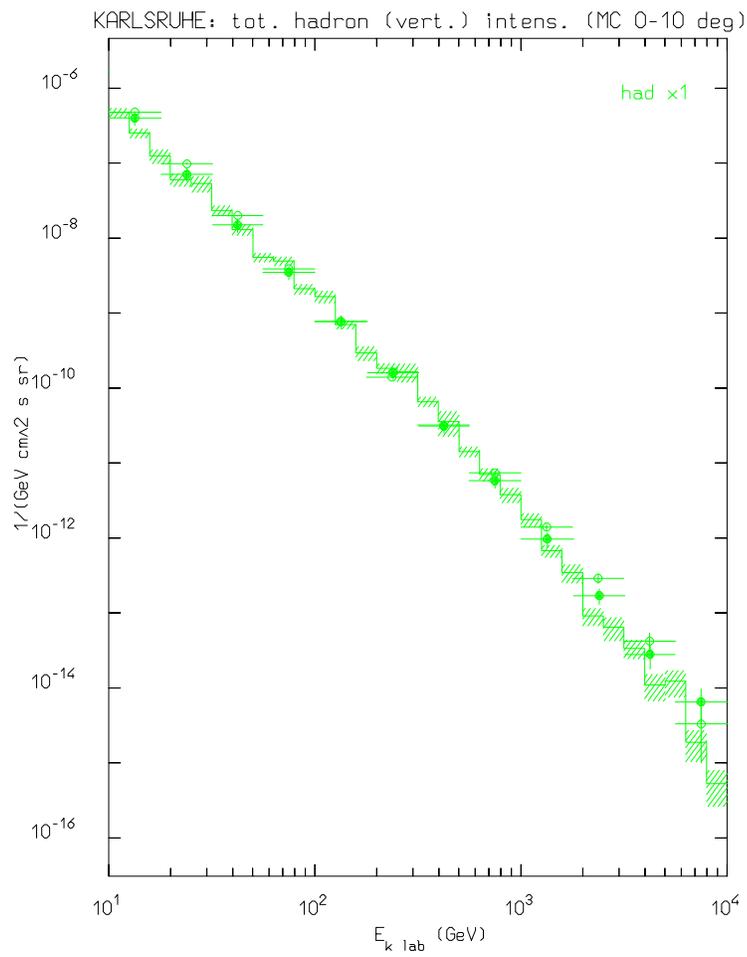,width=100mm%
,bbllx=60pt,bblly=100pt,bburx=590pt,bbury=740pt}}
\caption{Hadron flux measured with the KASKADE 
experiment~\protect\cite{KASKADE}.
Histogram is simulation result.
\label{fig:atmhad}
}
\end{center}
\end{figure}

The shower simulations in atmosphere have been compared also to the most 
recent hadron spectra at different
latitudes and altitudes, obtaining remarkable agreement. As an example,
in fig.~\ref{fig:atmhad} we compare MonteCarlo results to the 
hadron flux measured with the KASKADE experiment\cite{KASKADE}.

\clearpage
\section{Conclusions}
\label{sec:concl}
The phenomenological study of hadronic interactions is still a fundamental 
issue for astroparticle physics. Now that, at least for primary energy 
lower than 100 GeV, the uncertainties on primary spectrum have been
substantially reduced by the quality of new experiments like AMS\cite{ams} and
BESS\cite{bess}, the need for a better quality model of hadronic
interactions is even more necessary if accuracy of predictions has to be 
pursued. For this goal the ``microscopic'' codes are mostly recommended,
thanks to their predictive power in a very large kinematic region,
constrained by a limited number of parameters.
Uncertainties on the modelling of hadronic interactions will remain
a fundamental issue, and probably only new data,
if experimental systematics can be kept under reasonable control,
will help model builders. The HARP experiment\cite{harp} at CERN is aiming
at this goal. 
This kind of activity is beneficial not only for particle physics 
and astrophysics, but also for applied
science, since these calculations are necessary to 
understand radiation fluxes in the Earth's atmosphere, and this is
of great interest for civil aviation and for the design of 
satellite activities\cite{roesler}.
The \FLUKA{} MonteCarlo model is already being used for this purpose:
doses to commercial flight are the subject of a work in progress, together
with the development of a specific model for heavy ion transport and
interaction: this will be of the utmost importance for
dose and damage calculation in space aircrafts.

\end{document}